\documentclass{article}
\usepackage{spconf,amsmath,graphicx}
\usepackage{url}
\title{Multi-task Learning for Speaker Verification and Voice Trigger Detection}
\name{Siddharth Sigtia, Erik Marchi, Sachin Kajarekar, Devang Naik, John Bridle}
\address{Apple}
\begin{document}
%\ninept
%
\maketitle
\begin{abstract}

Automatic speech transcription and speaker recognition are usually treated as separate tasks even though they are interdependent. In this study, we investigate training a single network to perform both tasks jointly. We train the network in a supervised multi-task learning setup, where the speech transcription branch of the network is trained to minimise a phonetic connectionist temporal classification (CTC) loss while the speaker recognition branch of the network is trained to label the input sequence with the correct label for the speaker. We present a large-scale empirical study where the model is trained using several thousand hours of labelled training data for each task. We evaluate the speech transcription branch of the network on a voice trigger detection task while the speaker recognition branch is evaluated on a speaker verification task. Results demonstrate that the network is able to encode both phonetic \emph{and} speaker information in its learnt representations while yielding accuracies at least as good as the baseline models for each task, with the same number of parameters as the independent models. 

\end{abstract}
\section{Introduction}
\label{sec:intro}

Speech based personal assistants allow users to interact with devices like phones, watches, speakers, and headphones via speech commands. Usually the speech commands are prefixed with a \emph{trigger phrase}. Therefore, accurately detecting the trigger phrase is important as it signals the start of a user interaction with a device. Detecting a given phrase involves 2 steps. The first is to decide if the phonetic content in the input audio matches that of the trigger phrase. This process is known as voice trigger detection \cite{MLBlogHS,sigtia2018}. The second is to determine whether the speaker's voice matches the voice of the registered user(s) of the device. This problem is known as speaker verification \cite{Marchi18-GDT}. 

Currently, these 2 problems are often considered independently. Voice trigger detection, which is interchangeably known as keyword spotting \cite{chen2014small}, wake-up word detection \cite{kumatani2017direct}, or hotword detection \cite{huang2019multi}, is treated as an acoustic modelling problem. The inputs to these models are the acoustic signal and they are trained to either produce a sequence of phonetic labels or to output binary labels indicating the presence or absence of a given trigger phrase. Recent approaches to this problem have explored a variety of neural network architectures \cite{sigtia2018,chen2014small,sainath2015convolutional,fernandez2007application}. Their primary aim is to recognise the phonetic content (or the trigger phrase directly) in the input audio, with \emph{no regard} for the identity of the speaker. 

On the other hand, speaker verification systems aim to confirm the identity of the speaker by comparing an input utterance with a set of \emph{enrolment utterances} which are collected when a user sets up their device. This task is done by learning a fixed-dimensional representation or \emph{embedding} that encodes information only related to the characteristics of the speaker while remaining invariant to the phonetic content of the audio. Given a test recording, the embedding for this recording is compared against the embeddings generated from the enrolment utterances using a suitable distance metric. Speaker verification algorithms can be characterised based on whether the phonetic content in the inputs is limited, which is known as text-dependent speaker verification \cite{heigold2016end}. Alternatively, text-independent systems operate with no restrictions on the phonetic content \cite{Marchi18-GDT}. Speaker verification systems are also classified according to the objective function used to train the embedding function. One common approach to learning a speaker embedding is to train a neural network to output correct speaker labels given the input \cite{Marchi18-GDT,heigold2016end,liu2019large}. An alternative strategy is to use the triplet loss \cite{li2017deep,wan2018generalized}, where the objective more explicitly encodes the notion that embeddings from the same speaker must be close while embeddings from different speakers should be far apart.   

Although these 2 tasks are related, they are treated independently when considering them as engineering problems. We believe that knowledge of the speaker would help determine the phonetic content in the acoustic signal and vice versa, therefore estimating both properties is similar to solving simultaneous equations. In this study, the main research question we try to answer is ``can a single network efficiently represent both phonetic and speaker specific information?". Rather than trying to estimate or exploit the interdependence between the two tasks, we explore whether using a shared/joint network to solve both tasks results in positive inductive transfer between them. From a practical standpoint, being able to share computation between the two tasks can save on-device memory, computation time or latency and the amount of power/battery consumed. More generally, we are interested in studying whether a single model can perform multiple speech understanding tasks rather than designing a separate model for each task.  

%is extremely useful since current systems are run on parallel threads and operate on different acoustic features, which is a limiting factor on mobile devices with limited memory and compute. 

This problem has received some attention in the literature recently. \cite{tang2016multi} proposes a recurrent architecture that tries to explicitly encode the interdependence between the phonetic and speaker recognition branches of the model. \cite{shafey2019joint} proposes a sequence-to-sequence model for performing joint speaker diarization and ASR for a limited number of speakers. \cite{adi2019reverse} investigates the effect of adding or removing speaker information while training a speech transcription model, though the limited size of the dataset prevents thorough evaluation on a speaker verification task. In this study, we train a joint network to perform a phonetic labelling task and a speaker recognition task. Our \textbf{main contribution} is to perform a large-scale empirical study where the joint model is trained using over 15,000 hours of labelled training data. We evaluate the 2 branches of the model on a voice trigger detection task and a speaker verification task, respectively. The models are compared against strong baseline models on challenging real-world evaluation sets. The results presented demonstrate that it is possible for a single network to encode both speaker and phonetic information and yield similar accuracies as the baseline models \emph{without} requiring any additional parameters. %This signals that it should be possible to design more efficient architectures where a single on-device model performs multiple speech understanding tasks rather than a different model for each task. 

%We train the joint network on large-scale datasets with thousands of hours labelled training data. Unlike some of the studies mentioned before, we evaluate the performance of the joint model on \emph{both} tasks on challenging test sets in a variety of conditions. Our main contribution is to demonstrate that it is possible for a single network to yield the same accuracies on both tasks as baseline models trained separately for each task. We show that the joint model is able to do this without requiring any additional parameters. 

\section{Voice Trigger Detection Baseline}

The baseline architecture for the voice trigger detector is as follows: we extract 40-dimensional log-filterbanks from the audio at 100 frame-per-second (FPS). At every step, 7 frames are spliced together to form symmetric windows and finally this sequence of windows is sub-sampled by a factor of 3, yielding a 280-dimensional input vector to the model at a rate of 33 FPS. The features are input to a stack of 4 bidirectional LSTM layers with 256 units in each layer (Figure 1). This is followed by an output softmax layer over context-independent phonemes and additional sentence and word boundary symbols, resulting in a total of 53 output symbols. This model is then trained by minimising the CTC loss function \cite{graves2006connectionist}. Note that at inference, we want to use the model to compute the probability of the trigger phrase phone sequence given the acoustic evidence, $P(\text{TriggerPhrasePhoneSeq}|\mathbf x)$. This computation can be compactly expressed as a left-to-right HMM and consequently the probability scores can be efficiently computed using dynamic programming. The main attraction of using this setup is the fact that we can use the same training data as the main ASR without requiring a separate training dataset specific to each trigger phrase. 

The training data for this model is 5000 hours of anonymised audio data that is manually transcribed, where all of the recordings are sampled from intentional voice assistant invocations and are assumed to be \emph{near-field}. A third of the training examples are additionally convolved with room impulse responses (RIRs) to simulate reverberant speech. We use a set of 3000 different RIRs that are internally collected in a wide range of houses and represent a diverse set of acoustic conditions. Furthermore, a third of the data is also mixed with echo residuals to simulate playback from the device at various levels \cite{MLBlogFrontEnd}. The model parameters are optimised using large-batch stochastic gradient descent (SGD). Each mini-batch contains 128 utterances and we use 32 GPUs in parallel resulting in an effective batch size of 4096 examples per gradient update. We use an initial learning rate of 0.0001 and update the weights using the Adam optimiser. 

\section{Speaker Verification Baseline}

The inputs to this model are exactly the same as the inputs to the model described above. The baseline model comprises 2 bidirectional LSTM layers with 256 units each. Rather than using the activations of the final LSTM hidden state as the speaker embedding as in \cite{Marchi18-GDT}, we use a simple location-based attention mechanism \cite{luong2015effective} to summarise the encoder activations as a fixed-dimensional vector. We found the attention mechanism to be particularly effective in the text-independent setting (c.f. Section 5). Let the activations of the final layer of the encoder be $\mathbf h = (\mathbf h_1,\ldots,\mathbf h_{T})$ where $\mathbf{h}_t$ is a 512-dimensional vector and represents the encoder activations at time-step $t$ obtained by concatenating the 256-dimensional activations from the final forward and backward LSTM layers. At each time-step, we compute a scalar valued score: 

\begin{equation}
	s_t = f_{attn}(\mathbf{h}_t,\boldsymbol \theta_{attn}),
\end{equation}
where $f_{attn}$ is an MLP with 1 hidden layer with 256 units and a scalar output $s_t$, $\boldsymbol \theta_{attn}$ are the weights and biases of the MLP. The scores at each time-step are normalised: 
 
\begin{equation}
	\alpha_t = \frac{\exp(s_t)}{\sum_t \exp(s_t)},
\end{equation}
and the final summary vector is obtained by computing a weighted sum of encoder activations:

\begin{equation}
	\mathbf{e} = \sum_t \alpha_t \mathbf{h}_t,
\end{equation}
where $\mathbf{e}$ is a 512-dimensional vector. The speaker embedding is obtained by applying a 128-dimensional linear projection to the vector $\mathbf{e}$. During training, the embedding layer is followed by a softmax layer over the number of speakers in the training dataset and the network is trained by minimising the categorical cross-entropy loss. 
%Therefore, the underlying assumption is that training in such a manner would result in an embedding space that is able to separate out the characteristics that are unique to each speaker. 

\begin{figure}[t]
\begin{minipage}[]{1.0\linewidth}
  \centering
  \centerline{\includegraphics[width=\textwidth]{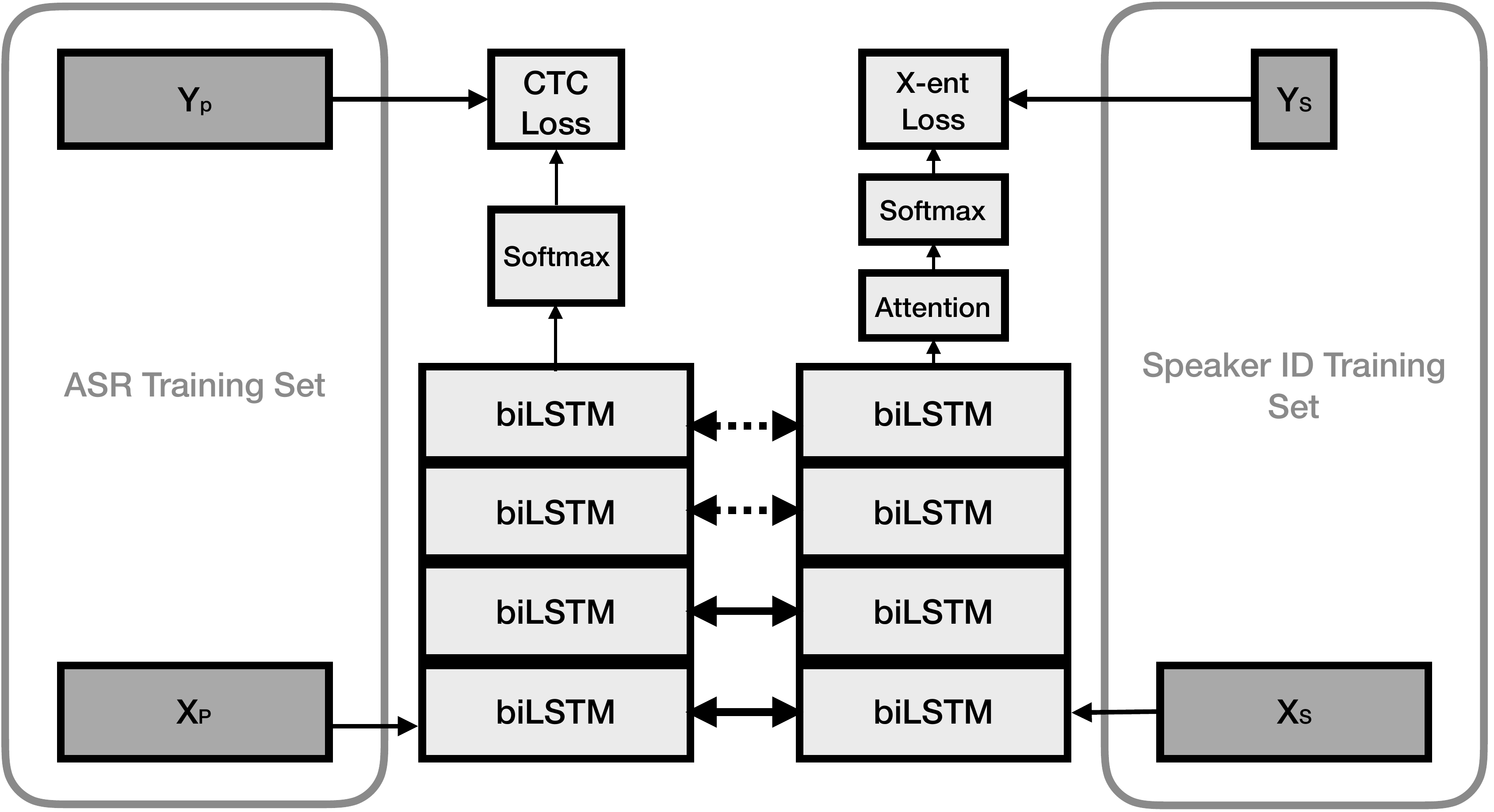}}
  \caption{The left branch of the model represents the voice trigger detector, the right branch is the speaker verification model. Solid horizontal arrows represent layers with tied weights, dashed arrows represent layers with weights that may or may not be tied. }
\end{minipage}
\end{figure}
%\vspace{-8mm}

During inference, given a test utterance $\mathbf{x}$, the speaker embedding is obtained by removing the final softmax layer and using the 128-dimensional activations of the previous layer. A score for the test utterance is then obtained by computing cosine similarities between $\mathbf{x}$ and the enrolment utterances:

%\vspace{-6mm}
\begin{equation}
	s(\mathbf{x},spk) = \frac {1}{N} \sum_{i=1}^{N} \frac{f(\mathbf{x})^T f(\mathbf{x}_{i}^{spk})}{\| f(\mathbf{x})\| \| f(\mathbf{x}_{i}^{spk})\| },
\end{equation}
where $f$ denotes the embedding function, $\mathbf{x}$ is a test input, $spk$ is the identifier for a given speaker and $\mathbf{x}_{i}^{spk}$ denotes the $i^{\text{th}}$ enrolment utterance for that speaker. Finally, a decision is made to accept or reject the test utterance by comparing the score against a threshold. 

The training data for the speaker recognition task comprises 4.5 million utterances sampled from intentional voice assistant invocations. The training set contains 21,000 different speakers, with a minimum of 20 examples and a median of 118 examples per speaker, resulting in a training set with over 5700 hours of audio. Note that the training labels only contain the labels of the speaker \emph{without} any information about the phonetic content in the audio. Each training utterance is of the form ``Trigger phrase, payload" for e.g.``Hey Siri (HS), play me something I'd like". For every training example, we generate 3 segments: the trigger phrase, the payload and the whole utterance. We found that breaking the utterances up this way results in models that generalise significantly better. The final dataset contains 13 million training examples with over 11,000 hours of labelled training data. We use exactly the same hyperparameters as the voice trigger baseline for the optimiser. 
%The research question that this study tries to answer is whether it is possible for a single network to encode both speaker and phonetic information in its learnt internal representations. While from a practical standpoint, it is extremely useful to be able to share computation while performing both tasks compared to running 2 parallel networks on different threads operating on different sets of features. This is of particular importance on mobile devices like phones, watches and headphones. 

\section{Multi-task Learning}

Figure 1 provides an overview of the multi-task learning (MTL) setup. Note that most of the weights in the two baseline systems are in modules with the same structure (biLSTM layers). To perform MTL, we share (tie) correspondong weights in some of those modules. The objective function that is used to optimise the parameters of the joint model is as follows:

%\vspace{-6mm}
\begin{equation}
	C_{mtl}{\big(\boldsymbol \theta_{tied},\boldsymbol \theta_{vt},\boldsymbol \theta_{spk} \big)} = C_{vt}{\big(\boldsymbol \theta_{tied},\boldsymbol \theta_{vt} \big)} + C_{spk}{\big(\boldsymbol \theta_{tied},\boldsymbol \theta_{spk} \big)},
\end{equation}
where $C_{vt}$ denotes the CTC loss for the voice trigger branch, $C_{spk}$ is the cross-entropy loss for the speaker recognition task, $\boldsymbol \theta_{tied}$ are the weights of the \emph{tied} biLSTM layers, $\boldsymbol \theta_{vt}$ are the \emph{untied} parameters in the voice trigger branch and $\boldsymbol \theta_{spk}$ are the parameters of the speaker verification branch of the model.

We train 3 sets of models. The first is where all 4 biLSTM layers of the encoder are tied for both tasks. This setup is restrictive since the model is expected to perform 2 different tasks with the same architecture and parameter count as the voice trigger baseline. Furthermore the model must also learn to represent both phonetic and speaker information using the \emph{same 512-dimensional activations} of the final layer of the encoder. The second model relaxes this constraint by sharing only 3 biLSTM layers in the encoder, with separate final biLSTM layers for the voice trigger and speaker recognition branches. Finally, we train a third model where 2 biLSTM layers have tied weights, with 2 additional biLSTM layers for each branch (Figure 1). Note that the total number of parameters in this third model is equal to \emph{the sum of the parameters} in the voice trigger and speaker baseline models. 

The training dataset for these models is formed by concatenating the datasets for the voice trigger and speaker recognition tasks. The resulting dataset contains over 16,000 hours of labelled training data, where 5000 hours of audio have phonetic labels and the remaining examples have speaker labels only. We shuffle the dataset after every training epoch, ensuring that each mini-batch contains examples for both tasks. We use exactly the same optimisation hyper-parameters as before. Somewhat surprisingly, training the joint model did not require any further hyper-parameter tuning. We observed that simply summing up the objective functions for both tasks with unity coefficients (Equation 4) resulted in a stable training objective. %Even though the scale of the individual costs differed by 2 orders of magnitude at the start of training, stochastic gradient descent was able to find parameters that minimised both objectives simultaneously. 

\section{Evaluation}

\begin{figure}[t]
  \begin{minipage}[b]{0.9\linewidth}
  \centering
  \centerline{\includegraphics[width=\textwidth]{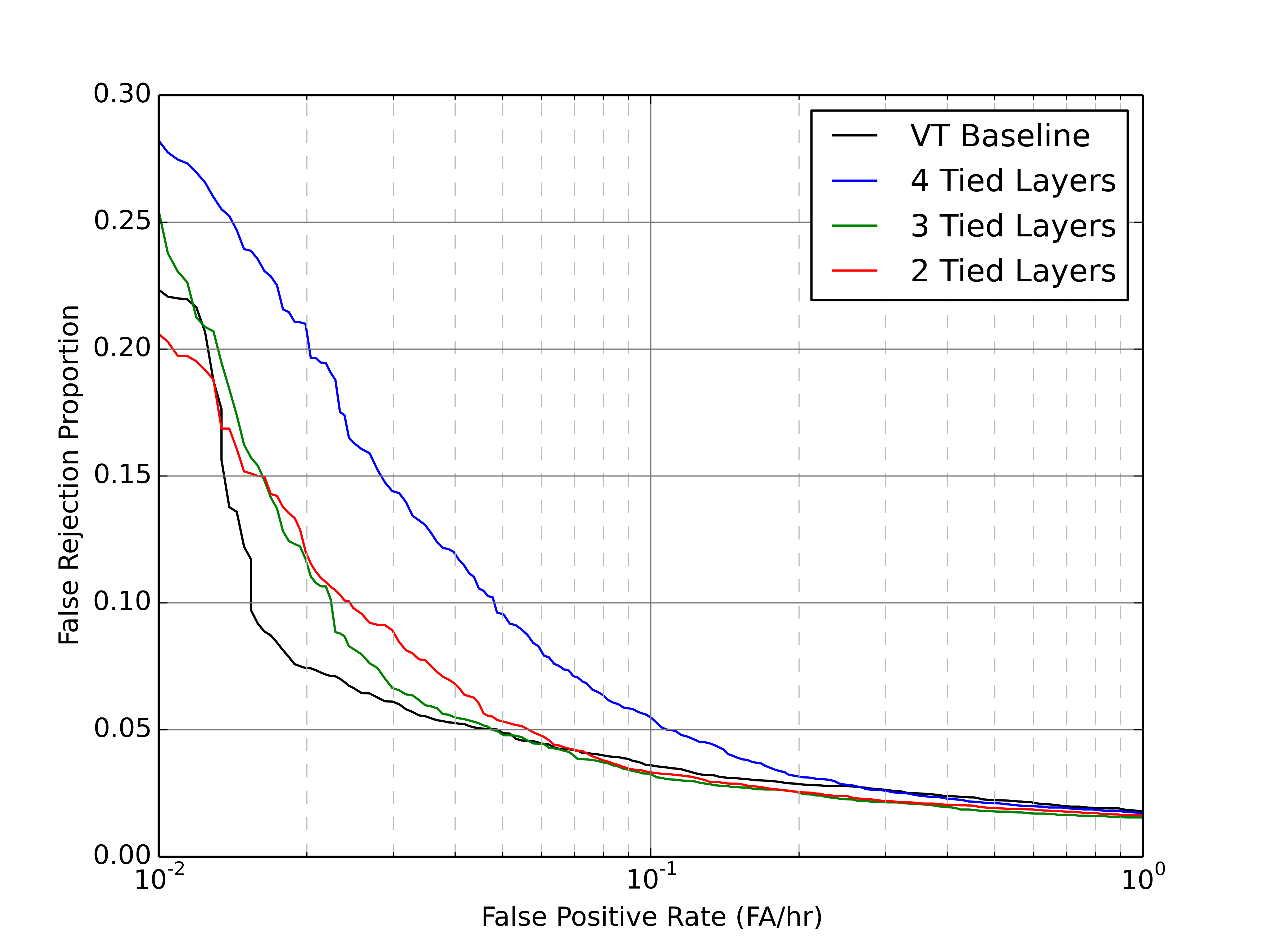}}
  \caption{DET curves for the voice trigger detection task.}
%  \vspace{2.0cm}
  %\centerline{(a) DET curves for the voice trigger detection task. }\medskip
\end{minipage}
  %\center
  %%\includegraphics[width=0.8\textwidth]{figures/det_curves.png}
  %\caption{DET curves show that the models trained with MTL yield significant improvements in accuracy in all acoustic conditions.}
  %\label{results}
\end{figure}

We evaluate the model described above on 2 tasks, voice trigger detection and speaker verification. Note that we employ a cascaded 2-stage architecture for the detection system \cite{MLBlogHS,sigtia2018}, where a low-power detector is always running and listening for the trigger phrase. If a detection is made at this stage, the acoustic segment is handed over to larger more complex models that verify both whether the segment contains the trigger phrase and the identity of the speaker. All the models discussed so far are used in this second pass. 
\begin{table}[t]
  \begin{center}
    \begin{tabular}{l|c|c|c} % <-- Alignments: 1st column left, 2nd middle and 3rd right, with vertical lines in between
      \textbf{Model} & \textbf{HS} & \textbf{HS+Payload} & \textbf{Payload}\\
      %$\alpha$ & $\beta$ & $\gamma$ \\
      \hline
      Baseline & 2.45 & 2.11 & 8.01\\
      4 Tied Layers & 2.98 & 2.73 & 9.87\\
      3 Tied Layers & \textbf{2.25} & 2.55 & 7.78\\
      2 Tied Layers & 2.35 & \textbf{1.98} & \textbf{7.40}\\
    \end{tabular}
    \caption{EERs for the speaker verification task.}
    \label{tab:table1}
  \end{center}
\end{table}

For the detection task, we use the voice trigger branch of the model to compute the probability $P(\text{TriggerPhrasePhoneSeq}|\mathbf{x})$ given an input utterance $\mathbf{x}$ and this score is compared to a threshold to accept or reject the hypothesis that the input contains the trigger phrase. The model is evaluated on a large internally-collected test-set used for evaluating models designed for smart speakers. All the data is recorded using the device in live sessions in realistic home environments. The test-set contains 100 subjects in total with an equal number of male and female participants, where each subject makes a series of prompted commands, all beginning with the trigger phrase. The subjects are at distances between 8 and 15 feet from the device. We collect over 13,000 such \emph{positive} utterances in 4 acoustic settings: (a) quiet room, (b) external noise from a TV or kitchen appliance in the room, (c) music playback from the recording device at medium volume, and (d) music playback from the recording device at loud volume.  Note that condition (d) is challenging due to high levels of residual noise that the voice-trigger model must contend with in order to detect the trigger-phrase. These examples are used to measure the proportion of utterances that are falsely rejected (FRs) by the system. In addition to these recordings, the test set also consists of 2,000 hours of continuous audio recordings from TV, radio, and podcasts which \emph{do not} contain the trigger phrase. This data allows the measurement of the false-alarm (FA) rate in terms of FA’s per hour of active external audio. The evaluation results are presented as modified detection-error trade-off curves to compare accuracy between models (Figure 2). Each curve displays the FA and FR rate while sweeping the trigger threshold for a particular model.

%The model is evaluated on a large test set that contains 12,000 utterances that start with the trigger phrase. These utterances are used to measure the percentage of examples that are \emph{false rejections} by the model. Additionally, we use a dataset of 10,000 hours of general speech data that \emph{do not} contain the trigger phrase to measure the rate of \emph{false accepts} by the system. The positive utterances with the trigger phrase are collected in a wide range of acoustic conditions like near and far-field, external noise sources, music playback and so on in order to simulate challenging real-world test conditions. The evaluation results are presented in the form of detection error trade-off curves, where the Y-axis represents the proportion of utterances that are false rejected and the log-scale X-axis measure the number of false-alarms per hour of active speech. Each point on the curve corresponds to a different operating threshold. The results are summarised in Figure 2. 

The evaluation dataset for the speaker verification task contains data from 500 speakers not heard during training, with 2500 enrolment utterances in total and a median of 5 enrolment utterances per speaker. The test data contains 53,000 utterances with a minimum of 40 test utterances per speaker and a median of 106 utterances per speaker. Note that like the training data, all the test utterances are assumed to be near-field. We evaluate the models in 3 different settings: the first is the text-dependent setting where only the trigger phrase portion of the utterance is used to compute the speaker embedding. The second setting corresponds to using the full utterance (HS+Payload), while the third setting corresponds to using only the payload to compute the embedding and represents the text-independent setting.  As is standard practice, we use the equal-error rate (EER) as a metric for comparing the models along with t-norm score normalisation \cite{auckenthaler2000score}. The results are summarised in Table 1. 

From Figure 2 we observe that the models with 2 (red) and 3 (green) tied layers yield roughly similar accuracies to the baseline model (black). The fully tied model (blue) yields notably worse accuracies compared to the baseline across all operating points. This result supports our earlier hypothesis that tying all layers and sharing the same activations to represent both phonetic and speaker information is very restrictive. For FA rates below 0.05 FAs/hour, the 2 and 3 tied layer models (red and green curves) yield marginally worse FR rates compared to the baseline, but for over the half the points the two tied models yield marginally better accuracies. On the speaker verification evaluations (Table 1), we observe that the model with 2 tied layers outperforms the baseline model in \emph{all 3 settings}. There is no significant difference on the easier text-dependent task, however on the text-independent task we observe a relative improvement of $7.6\%$ over the baseline. The model with 3 tied layers yields results at least as good as the baselines on both tasks (except for the HS+Payload case) with 1 LSTM layer less than the baselines. These results indicate that training the same model to perform phoneme transcription and speaker recognition can result in positive inductive transfer between tasks. Another interesting feature of these results is that the model was trained using \emph{disjoint} datasets i.e.\ each audio example has either phonetic \emph{or} speaker labels, never both. This observation suggests a flexible design where it is possible to train a model on multiple related tasks by concatenating training data for different tasks, rather than obtaining multiple labels for each training example. 

\section{Conclusions}

%We trained a single model with some shared layers to perform phonetic transcription and speaker recognition on a large labelled dataset. We evaluated the joint model on a speaker verification and a voice trigger detection task. Our results demonstrated that the model is capable of encoding both phonetic and speaker specific properties in its learnt representations. The proposed model yields accuracies at least as good as the baseline without requiring any additional parameters. 

%We trained a single model with some shared layers to perform phonetic transcription and speaker recognition on a large labelled dataset. 

Our results demonstrate that sharing the first two layers of the model between the speaker and phonetic tasks gives accuracies that are as good as the individual baselines. This result indicates that it is possible to share some of the low-level computation between speech processing tasks without hurting accuracies. We hope to train such unified models to perform a larger combination of speech understanding tasks while achieving positive inductive transfer between tasks. In future work, we plan to train models on a larger set of related speech tasks while exploiting their interdependence.  
%We investigated whether a single model can encode both phonetic and speaker specific information in its learnt representations. We trained a single model with some shared layers and some layers specific to each task on a large training set. We evaluated the 2 branches of the model on a speaker verification task and a voice trigger detection task. Our results demonstrate that the joint model yields similar results to the baseline models for each task without requiring any extra parameters. 
%\vfill\pagebreak
\bibliographystyle{IEEEbib}
\bibliography{strings,refs}

\end{document}